\documentclass[aps,prl,twocolumn,groupedaddress,superscriptaddress,floatfix]{revtex4}

\usepackage[utf8]{inputenc}
\usepackage[T1]{fontenc}
\usepackage{slashed}
\usepackage{times}

\usepackage{amssymb,amsfonts,amsmath,amsthm}

\usepackage{slashed}
\usepackage{dcolumn}
\newcolumntype{l}[1]{D{.}{\cdot}{#1}} 

\usepackage{graphicx}
\usepackage[tight,hang,raggedright,normalsize]{subfigure}

\usepackage[
            pdftitle={A possible explanation of the size of D mixing},
            pdfauthor={Alexander Lenz},
            pdfsubject={The size of D mixing},
            pdfkeywords={mxing of netural mesons, new physics},
            pdfpagemode=UseOutlines,
            colorlinks=true,
            pdfproducer={University of Regensburg},
            ]{hyperref}

\begin{document}


\title{$D-\bar D$ mixing in the framework of the HQE revisited}

\author{M. \surname{Bobrowski}}
    \affiliation{Institut f\"ur Theoretische Physik, Universit\"at Regensburg, 93040 Regensburg, Germany}
\author{A. \surname{Lenz}}
    \affiliation{Institut f\"ur Theoretische Physik, Universit\"at Regensburg, 93040 Regensburg, Germany}
\author{J. \surname{Riedl}}
    \affiliation{Institut f\"ur Theoretische Physik, Universit\"at Regensburg, 93040 Regensburg, Germany}
\author{J. \surname{Rohrwild}}
    \affiliation{Institut f\"ur Theoretische Physik, Universit\"at Regensburg, 93040 Regensburg, Germany}


\begin{abstract}
We reconsider the leading HQE contributions to the absorptive part of the mixing amplitude of neutral
$D$ mesons by taking also $\alpha_s$ corrections and subleading $1/m_c$ corrections into account.
We show that these contributions to $\Gamma_{12}$ do not vanish in the exact SU(3)$_F$ limit and 
they also can have a large phase. Moreover, we give an example of a new physics model that can 
enhance these leading HQE terms to $\Gamma_{12}$, which are orders of magnitude lower than the 
current experimental expectation, by a factor in the upper double-digit range.
\end{abstract}

\maketitle

\begin{figure}[t]
\includegraphics[width=0.22\textwidth,clip, angle=0]{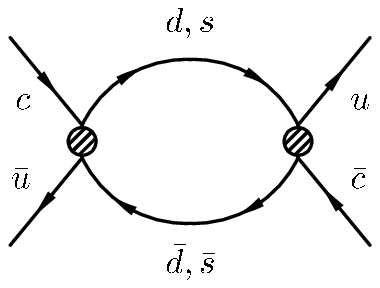}
\hfill
\includegraphics[width=0.22\textwidth,clip, angle=0]{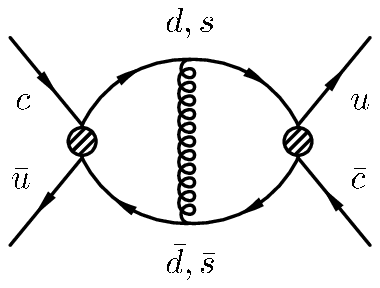}
\vspace*{-0.5cm}
\caption{Contributions to $\Gamma_{12}$ from operators of dimension 6 (D=6). The leading order 
QCD diagram is shown in the left panel, an example for $\alpha_s$ corrections is shown 
in the right panel.}
\label{fig:D=6}
\vspace*{-0.5cm}
\end{figure}

\phantomsection
\addcontentsline{toc}{subsubsection}{Introduction.}
\paragraph{Introduction. ---}
Mixing of neutral mesons provides an excellent testing ground for the standard model (SM) 
and its possible extensions. In the $D^0$ system the following quantities have been 
measured \cite{Barberio:2008fa}
\vspace*{-0.2cm}
\begin{equation}
y := \frac{\Delta \Gamma}{2 \Gamma_D} = \frac{7.3 \pm 1.8}{1000},
\hspace{0.1cm}
x :=  \frac{\Delta M}{\Gamma_D} = \frac{9.1^{+2.5}_{-2.6}}{1000}.
\end{equation}
\vspace*{-0.2cm}
\phantomsection
\addcontentsline{toc}{subsubsection}{Mixing formalism}
\paragraph{Mixing formalism. ---}
The mixing of neutral mesons is described by box diagrams with the absorptive part $\Gamma_{12}$
and the dispersive part $M_{12}$. The observable mass and decay rate differences are given
by ($ \phi := \arg [ - M_{12} / \Gamma_{12}  ]$)
\begin{eqnarray}
\left( \Delta M \right)^2 - \frac14 \left( \Delta \Gamma \right)^2
& = &
4 |M_{12}|^2 - |\Gamma_{12}|^2 ,
\nonumber \\
\Delta M \Delta \Gamma 
& = &
4 |M_{12}| |\Gamma_{12}| \cos (\phi) \, .
\end{eqnarray}
If $|\Gamma_{12}/M_{12}| \ll 1$, as in the case of the $B_s$ system ($ \approx 5 \cdot 10^{-3}$) or if
$\phi \ll 1$, one gets the famous approximate formulae
\begin{equation}
\Delta M = 2 |M_{12}|\, , \hspace{0.5cm} \Delta \Gamma = 2 |\Gamma_{12}| \cos \phi \, . 
\label{approx}
\end{equation}
The experimental values for $x$ and $y$ suggest that in the $D^0$ system 
$|\Gamma_{12}/M_{12}| \approx {\cal O} (1)$, the size of the mixing phase $\phi$ will be 
discussed below.
%
\phantomsection
\addcontentsline{toc}{subsubsection}{Leading HQE predictions}
\paragraph{Leading HQE predictions. ---}
The absorptive part of the box diagram with internal $s$ and $d$ quarks 
can be decomposed according to the CKM structure as
\begin{equation}
\Gamma_{12} = 
- \left( \lambda_s^2 \Gamma_{ss} 
      +2 \lambda_s \lambda_d \Gamma_{sd} 
       + \lambda_d^2 \Gamma_{dd} \right), 
\label{Gamma12}
\end{equation}
with $\lambda_x = V_{cx} V_{ux}^*$. 
The application of the heavy quark expansion (HQE), which turned out to be very successful in the 
$B$ system, to the charm system typically meets major doubts. 
Our strategy in this work is the following: instead of trying to clarify the convergence of the HQE
in the charm system in advance, we simply start with the leading term and determine corrections to it.
The size of these corrections will give us an estimate for the convergence of the HQE in the $D^0$ system.
To this end  we first investigate the contribution of dimension-6 (D=6) operators
to $\Gamma_{12}$, see Fig.~(\ref{fig:D=6}).
Next we include NLO-QCD corrections, which were calculated for the $B_s$ system 
\cite{Beneke:1998sy,Beneke:2003az,Ciuchini:2003ww,Lenz:2006hd} and 
subleading terms in the HQE (dimension-7 operators), which were obtained in 
\cite{Beneke:1996gn,Dighe:2001gc}.
To investigate the size of $\alpha_s$ and $1/m_c$ corrections in more detail we 
decompose $\Gamma_{ss,sd,dd}$ into the Wilson coefficients $G$ and $G_S$ of the
$\Delta C=2$ operators $Q$ and $Q_S$ (for more details see \cite{Lenz:2006hd})
\begin{equation}
\Gamma_{xx}^{D=6,7} =
 \frac{G_F^2 m_c^2}{24 \pi M_{D^0}} \left[
G^{xx} \langle D | Q | \bar{D} \rangle + G_S^{xx} \langle D | Q_S | \bar{D} \rangle \right] + \Gamma_{xx}^{\frac{1}{m_c}},
\label{oldbasis}
\end{equation}
with
\vspace{-1cm}
\begin{eqnarray}
Q   & = & \bar{u}_\alpha \gamma^\mu (1-\gamma_5) c_\alpha \cdot 
          \bar{u}_\beta  \gamma_\mu (1-\gamma_5) c_\beta,
\nonumber
\\ 
Q_S & =& \bar{u}_\alpha (1+\gamma_5) c_\alpha \cdot 
         \bar{u}_\beta  (1+\gamma_5) c_\beta.
\label{DeltaC=2op}
\vspace*{-0.3cm}
\end{eqnarray}
The effect of the QCD corrections has already been discussed in  \cite{Golowich:2005pt}. 
In our numerics we carefully expand in $\alpha_s$: the leading order QCD contribution consists of 
leading order $\Delta C=1$ Wilson coefficients inserted in the left diagram of Fig.~(\ref{fig:D=6}), 
while our NLO result consists of NLO  $\Delta C=1$ Wilson coefficients inserted in both diagrams of 
Fig.~(\ref{fig:D=6}) and consistently throwing away all terms which are explicitly of 
${\cal O}(\alpha_s^2)$. Following \cite{Beneke:2002rj}
we have also summed terms like $z \ln z$ to all orders, therefore we use in our numerics
$\bar{z} = \bar{m}_s(\bar{m}_c)^2 /  \bar{m}_c(\bar{m}_c)^2 \approx 0.0092$.
The matrix elements in Eq.~(\ref{DeltaC=2op}) are parameterized as
\vspace*{-0.2cm}
\begin{eqnarray}
\langle D | Q          | \bar{D} \rangle & = &   \frac{8}{3} f_D^2 M_D^2 B(\mu) \, ,
\\
\langle D | Q_S        | \bar{D} \rangle & = & - \frac{5}{3} f_D^2 M_D^2 
                       \left( \frac{M_D}{m_c(\mu) + m_u(\mu)} \right)^2 B_S(\mu) \, .
\nonumber
\vspace*{-0.2cm}
\end{eqnarray}
We take $f_D = 212(14) $ MeV from \cite{Lubicz:2008am} and we derive 
$ B(m_c) = 0.9\pm 0.1$ and $ B_S(m_c) \simeq 0.9$ from 
\cite{Becirevic,Gorbahn:2009pp}.
Finally we use the $\overline{MS}$ scheme for the charm mass, $\bar{m}_c(\bar{m}_c) = 1.27$ GeV. 
For clarity we show in the following only the results for the central values of our parameters,
the error estimates will be presented at the end of the next section. We obtain
\begin{displaymath}
\begin{array}{|c||c|c|c|}
\hline
              & \mbox{LO} & \mbox{NLO} & \Delta \mbox{NLO} / \mbox{LO}
\\
\hline \hline
G^{ss}        & 0.25^{+0.09}_{-0.06} & 0.37^{+0.18}_{-0.20}       &  + 48 \%
\\
G^{ds}        & 0.26^{+0.09}_{-0.06} & 0.39^{+0.19}_{-0.21}       &  + 49 \%
\\
G^{dd}        & 0.29^{+0.09}_{-0.06} & 0.42^{+0.19}_{-0.22}       &  + 50 \%
\\
\hline
G^{ss}_S      & 1.97^{+0.15}_{-0.29} & 1.34^{+0.19}_{-0.23}       &  - 32 \%
\\
G^{ss}_S      & 1.98^{+0.15}_{-0.29} & 1.34^{+0.19}_{-0.23}       &  - 31 \%
\\
G^{ss}_S      & 1.98^{+0.15}_{-0.29} & 1.35^{+0.19}_{-0.23}       &  - 32 \%
\\
\hline
\end{array}
\end{displaymath}
For the error estimate we vary $\mu_1$ between $1$~GeV and $2 m_c$.
Combining $G$ and $G_S$ to $\Gamma_{xx}$ we get 
(in units of $ps^{-1}$)
\begin{displaymath}
\begin{array}{|c||c|c|c|c|}
\hline
              & \mbox{LO} & \Delta \mbox{NLO-QCD} & \Delta 1/m_c & \sum
\\
\hline \hline
\Gamma_{ss}        & 2.68      & -0.68  (-25 \%)   & -0.76 (-28 \%)   & 1.24 
\\
\Gamma_{ds}        & 2.70      & -0.67  (-25 \%)   & -0.76 (-28 \%)   & 1.26
\\
\Gamma_{dd}        & 2.71      & -0.66  (-24 \%)   & -0.76 (-28 \%)   & 1.29
\\
\hline
\end{array}
\end{displaymath}
Using instead the operator basis suggested in \cite{Lenz:2006hd} with
\begin{eqnarray}
\tilde Q_S & =& \bar{u}_\alpha (1+\gamma_5) c_\beta \cdot 
         \bar{u}_\beta  (1+\gamma_5) c_\alpha \, ,
\\
\langle D | \tilde Q_S | \bar{D} \rangle & = &   \frac{1}{3} f_D^2 M_D^2  
                       \left( \frac{M_D}{m_c(\mu) + m_u(\mu)} \right)^2 \tilde B_S(\mu) \, ,
\nonumber
\end{eqnarray} 
and $ \tilde B_S(m_c) \simeq 1.1$  leads to
\begin{displaymath}
\begin{array}{|c||c|c|c|c|}
\hline
              & \mbox{LO} & \Delta \mbox{NLO-QCD} & \Delta 1/m_c & \sum
\\
\hline \hline
\Gamma_{ss}        & 1.71      & +0.02  (+1 \%)   & -0.34 (-20 \%)   & 1.40
\\
\Gamma_{ds}        & 1.73      & +0.04  (+2 \%)   & -0.34 (-20 \%)   & 1.42
\\
\Gamma_{dd}        & 1.74      & +0.05  (+3 \%)   & -0.34 (-20 \%)   & 1.45
\\
\hline
\end{array}
\end{displaymath}
All in all we get large QCD (up to $50\%$) and large 1/$m_c$ corrections (up to $30\%$)
to the leading D=6 term, which considerably lower the LO values.
Despite these large corrections, the HQE seems not to be completely off. From 
our above investigations we see no hints for a breakdown of OPE. 
The same argument can be obtained  from the comparison of B and D meson lifetimes. 
In the HQE one obtains
\begin{equation}
\frac{\tau_B}{\tau_D} =
\frac{\Gamma_{0,D} + \delta \Gamma_D}{\Gamma_{0,B} + \delta \Gamma_B}
\approx  \frac{\Gamma_{0,D}}{\Gamma_{0,B}} \left( 1 + \frac{\delta \Gamma_D}{\Gamma_{0,D}} \right) 
                                     \left( 1 - \frac{\delta \Gamma_B}{\Gamma_{0,B}} \right)\;,
\end{equation}
where the leading term $\Gamma_0 \propto m_{b,c}^5 V_{CKM}^2$ corresponds to the free quark 
decay and all higher terms in the HQE are comprised in $\delta \Gamma$. 
For the ratio $\Gamma_{0,D} / \Gamma_{0,B}$ one gets a value close to one. 
Higher order HQE corrections in the B system are known to be smaller than 10 \% \cite{Lenz:2008xt}.
Using the experimental values for the lifetimes we get
\vspace{-0.3cm}
\begin{eqnarray}
\frac{\tau_B}{\tau_D} & \approx &  1.4 ... 4 \, (\mbox{Exp.}) 
\approx  1 \cdot \left( 1 + \frac{\delta \Gamma_D}{\Gamma_{0,D}} \right)\;.
\end{eqnarray}
From this rough estimate one expects higher order HQE corrections in the
D system of up to 300 \%. So clearly no precision determination will  
be possible within the HQE, but the estimates should still be within
the right order of magnitude.
%
%
%
%
\phantomsection
\addcontentsline{toc}{subsubsection}{Cancellations}
\paragraph{Cancellations. ---} 
As is well known huge GIM cancellations 
\cite{GIM} arise in the leading HQE terms for $D$ mixing. To make these effects more obvious, 
we use the unitarity  of the CKM matrix ($\lambda_d + \lambda_s + \lambda_b = 0$) to rewrite the expression
for the absorptive part in Eq.~(\ref{Gamma12}) as 
\begin{equation}
\Gamma_{12} = - \lambda_s^2 \left( \Gamma_{ss}\! - 2 \Gamma_{sd} + \Gamma_{dd}  \right)
+2 \lambda_s \lambda_b \left( \Gamma_{sd} - \Gamma_{dd} \right)
- \lambda_b^2 \Gamma_{dd} 
\label{cancel}
\end{equation}
Note that the CKM structures differ enormously in their numerical values: 
$\lambda_{d,s} \propto \lambda$ and $\lambda_b \propto \lambda^5$ 
\cite{Vub}
in terms of the Wolfenstein parameter $\lambda \approx 0.2255$. 
In the limit of exact $SU(3)_F$ symmetry, 
$\Gamma_{ss} = \Gamma_{sd} = \Gamma_{dd}$ holds and therefore, contrary to many
statements in the literature,
$\Gamma_{12} = - \lambda_b^2 \Gamma_{dd}$ is not zero but strongly CKM suppressed.
%
%
Next we expand the arising terms in $\bar{z}$. 
Using Eq.~(\ref{oldbasis}) we get in LO
\begin{eqnarray}
\Gamma_{ss}^{D=6} & = & 2.71252 - 3.14543 \bar{z} - 9.98427 \bar{z}^2 + ... \, ,
\nonumber
\\
\Gamma_{sd}^{D=6} & = & 2.71252 - 1.57271 \bar{z} - 4.99214 \bar{z}^2 + ... \, .
\end{eqnarray}
The first term in the above equations obviously corresponds to $\Gamma_{dd}^{D=6}$.
For the combinations in Eq.~(\ref{cancel}) we get
\begin{eqnarray}
\left( \Gamma_{ss} - 2 \Gamma_{sd} + \Gamma_{dd} \right)^{D=6} & = & -26.1794 \bar{z}^3 \approx \lambda^{7.3},
\nonumber
\\
\left( \Gamma_{sd} - \Gamma_{dd} \right)^{D=6} & = &  -1.57271 \bar{z} \approx \lambda^{2.8}.
\end{eqnarray}
To make the comparison with the arising CKM structures more obvious, we have expressed the size of these 
combinations also in terms of powers of the Wolfenstein parameter $\lambda$.
As it is well known, we find in the first term of Eq.~(\ref{cancel}) an extremly effective GIM cancellation,
only terms of order $\bar z^3$ survive.
In NLO we get
\begin{eqnarray}
\Gamma_{ss}^{D=6,7} & = & 1.2903 - 5.35346 \bar{z} - 8.77624 \bar{z}^2 + ... \, ,
\nonumber
\\
\Gamma_{sd}^{D=6,7} & = & 1.2903 - 2.67673 \bar{z} - 4.68377  \bar{z}^2 + ... \, .
\end{eqnarray}
The arising combinations in Eq.~(\ref{cancel}) read now
\begin{eqnarray}
\left( \Gamma_{ss} - 2 \Gamma_{sd} + \Gamma_{dd} \right)^{D=6,7} & = & 
0.59 \bar{z}^2 - 43.6  \bar{z}^3 \approx \lambda^{6.7} -  \lambda^{6.9},
\nonumber
\\
\left( \Gamma_{sd} - \Gamma_{dd} \right)^{D=6,7} & = &  -2.68 \bar{z} \approx \lambda^{2.5}.
\end{eqnarray}
The fact that now the first term of Eq.~(\ref{cancel}) is of order  $\bar z^2$ compared to 
$\bar z^3$ in the case of the LO-QCD value was discussed in detail in \cite{Beneke:2003az} and later on 
confirmed in \cite{Golowich:2005pt}. 
These numbers are now combined with  CKM structures, whose exact values read
\begin{eqnarray}
\lambda_d & = &  - c_{12} c_{23} c_{13} s_{12}   - c_{12}^2 c_{13} s_{23} s_{13} e^{i \delta_{13}}
                 = {\cal O} \left( \lambda^1 + i \lambda^5 \right),
\nonumber \\
\lambda_s & = &  + c_{12} c_{23} c_{13} s_{12} - s_{12}^2 c_{13} s_{23} s_{13} e^{i \delta_{13}} 
                 = {\cal O} \left( \lambda^1 + i \lambda^7 \right),
\nonumber \\
\lambda_b & = &  c_{13} s_{23} s_{13} e^{i \delta_{13}} = {\cal O} \left(\lambda^5 + i \lambda^5 \right),
\label{CKM3exact}
\end{eqnarray}
with $c_{ij} = \cos (\theta_{ij})$ and $s_{ij} = \sin (\theta_{ij})$.
Looking at Eq.~(\ref{CKM3exact}), it is of course tempting to throw away the small 
imaginary parts of $\lambda_d$ and $\lambda_s$, but we will show below that this is not
justified. Doing so and keeping only the leading term in the CKM structure
($c_{12} c_{23} c_{13} s_{12}$), which is equivalent to approximate $\lambda_b = 0$,
one gets a real $\Gamma_{12}$ which vanishes in the exact $SU(3)_F$ limit.
Keeping the exact expressions, we see that the first term in Eq.~(\ref{cancel}) is leading in CKM 
(${\cal O} \left[ \lambda^2 + i \lambda^8 \right]$) and has a negligible imaginary part, but it is suppressed 
by $ 0.6 \bar{z}^2 \approx \lambda^{6.7}$. The second term in Eq.~(\ref{cancel}) 
is subleading in CKM  (${\cal O} \left[ \lambda^6 + i \lambda^6 \right]$
and it can have a sizeable phase. This term is less suppressed by $SU(3)_F$ breaking 
($\approx 2.7  \bar z \approx \lambda^{2.5}$). 
The third term in Eq.~(\ref{cancel}) is not suppressed at all by  $SU(3)_F$ breaking, 
but it is strongly CKM suppressed  (${\cal O} \left[ \lambda^{10} + i \lambda^{10} \right]$). 
For clarity we compare the different contributions of Eq.~(\ref{cancel}) in the following table
\begin{displaymath}
\begin{array}{|c||c|c|c|}
\hline
           & \mbox{1st term}               & \mbox{2nd term}                     & \mbox{3rd term}
\\
\hline \hline
\mbox{LO}  & 26 \lambda_s^2 \bar z^3  &  3.1 \lambda_s \lambda_b \bar z & 2.7 \lambda_b^2
\\
           & \approx \lambda^{9.3}     & \approx \lambda^{8.4}          & \approx \lambda^{9.3} 
\\
\hline
\mbox{NLO} & 0.59 \lambda_s^2 \bar z^2  &  5.4 \lambda_s \lambda_b \bar z & 0.66 \lambda_b^2
\\
           & \approx \lambda^{8.7}     & \approx \lambda^{8.0}          & \approx \lambda^{10.3} 
\\
\hline
\end{array}
\end{displaymath}
From this simple power counting, we see that a priori no contribution to  Eq.~(\ref{cancel})
can be neglected. Having the comment \cite{Vub} in mind, 
we find however, that the first two terms of  Eq.~(\ref{cancel}) are of similar size, while
the third term is suppressed. Moreover, the second term can give rise to a large phase 
in $\Gamma_{12}$, while the first term has only a negligible phase.
To make our arguments more solid we perform the full numerics  
using the CKM values from \cite{CKM-Fitter} 
and obtain for the three contributions 
of  Eq.~(\ref{cancel}) 
\begin{displaymath}
\begin{array}{cccccc}
10^7 \Gamma_{12}^{D=6,7} & = &  -0.38 & + &  0.00002 i & (\mbox{1st term})
\\
                         &   & -6.34& - &  15.0   i & (\mbox{2nd term}) 
\\
                         &   & +0.19& - &  0.19   i & (\mbox{3rd term})
\\
                         & = & -6.53& - & 15.2    i & \! \! \! \! \! \!= (11 ... 38) \,  e^{-i (0.5... 2.3)} \, .
\end{array}
\end{displaymath}
Here we show for the first time the errors, they are estimated by varying $\mu_1$ between $1\;\text{GeV}$ and $2m_c$ and the 
dependence on the choice of the operator basis. 
The first term in Eq. (\ref{cancel}) turns out to be very sensitive with 
respect to the the exact values of the bag parameters and its real part can, within errors, be
numerically of the same size as the second term, which features a large imaginary part.
Furthermore, even the third term can give a non-negligible contribution, 
in particular to the imaginary part.

\begin{figure}[t]
\includegraphics[width=0.22\textwidth,clip, angle=0]{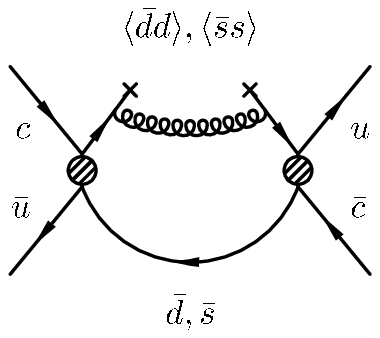}
\hfill
\includegraphics[width=0.22\textwidth,clip, angle=0]{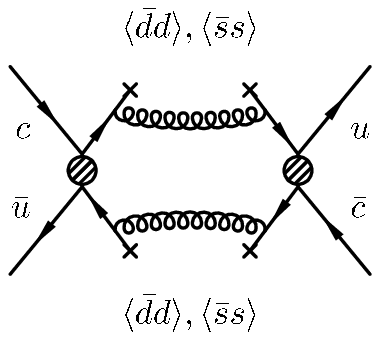}
\caption{Contributions to $\Gamma_{12}$ from operators of dimension 9 (D=9, left panel) and dimension
12 (D=12, right panel). To obtain an imaginary part the D=9 diagrams have to be dressed with 
at least one gluon and the D=12 diagrams with at least 2 gluons.}
\label{fig:D=9}
\end{figure}

\begin{figure}[t]
\vspace*{-1cm}
\hspace*{-.1\textwidth}\nolinebreak\includegraphics[width=.48\textwidth,clip, angle=270]{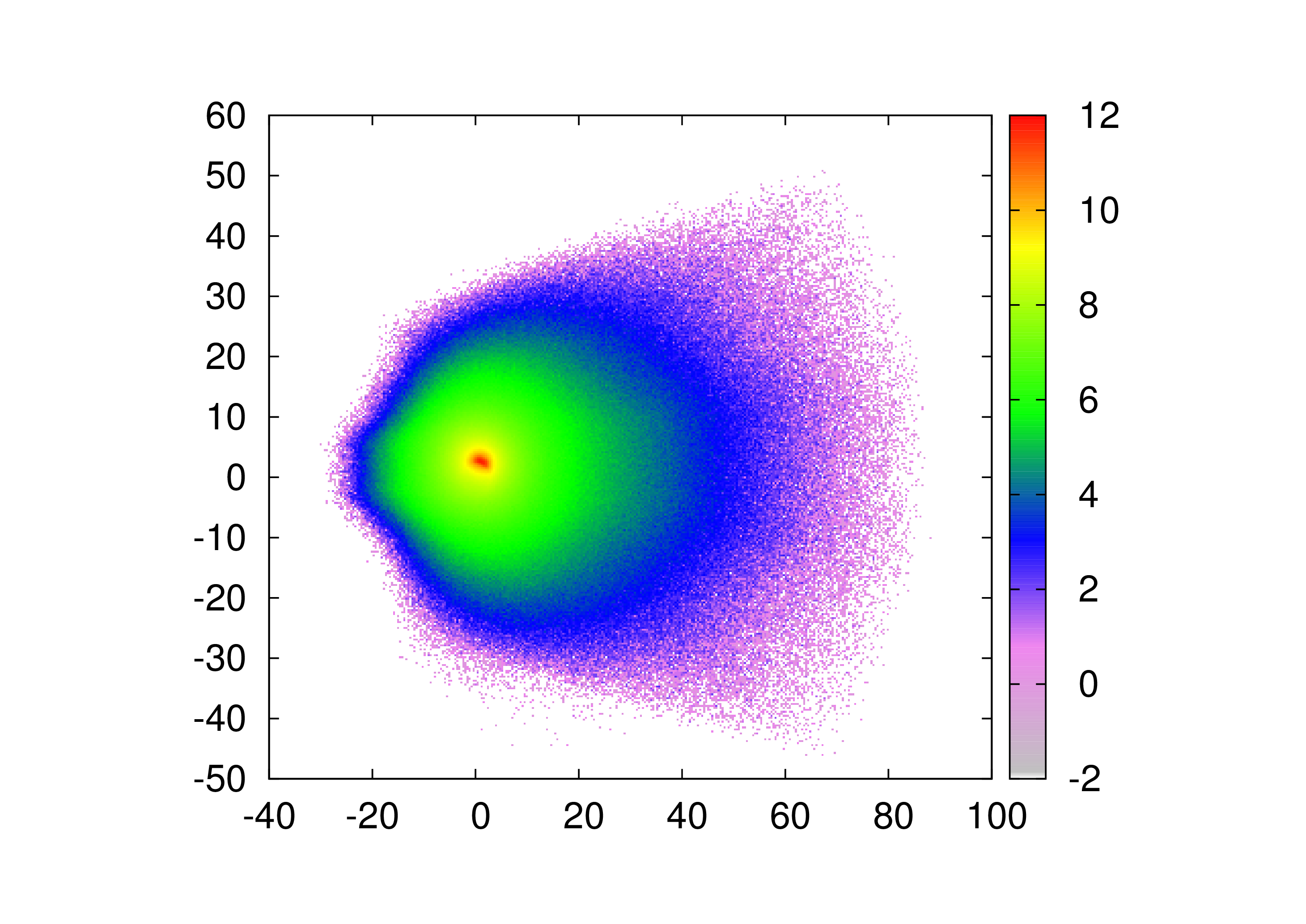}
\vspace*{-1cm}
\caption{The enhancement factor $\Gamma_{12}^{\text{SM4}}/\Gamma_{12}^{\text{SM3}}$ using the possible values for 
$V_{CKM4}$ found in \cite{Bobrowski:2009ng}. The color encoded scale denotes the logarithm of the number 
of allowed parameter points of $V_{CKM4}$.}
\label{fig:4gen}
\end{figure}

To summarize, we have demonstrated that the typical approximation $\lambda_b \approx 0$,
which is equivalent to neglecting the imaginary parts of $\lambda_d$ and $\lambda_s$ is wrong for
the case of the leading HQE prediction for $y_D$ and yields the wrong conclusion that 
$\Gamma_{12}$ cannot have a sizeable phase.
We get for the first terms in the OPE a value for $y_D$ of 
\begin{eqnarray}
y_D^{D=6,7;NLO} \leq |\Gamma_{12}| \cdot \tau_D & = & 4.6 \cdot 10^{-7} \! \! \! ... \, 1.6 \cdot 10^{-6}
\, . \, \, \, \, \, \, 
\end{eqnarray}
These values are still a factor of 
$0.8 ... 1.3 \cdot 10^4$ smaller than the experimental number.
This is in contrast to our previous expectations that the HQE should give at least the right 
order of magnitude. Moreover, we do not confirm the observation made in \cite{Golowich:2005pt} 
that the NLO result for $\Gamma_{12}$ is almost an order of magnitude larger than the LO result.
%
%
\phantomsection
\addcontentsline{toc}{subsubsection}{Higher HQE predictions}
\paragraph{Higher HQE predictions. ---}
In \cite{Georgi:1992as,Ohl:1992sr,Bigi:2000wn} higher order terms in the HQE of $D$ mixing 
were discussed. If the GIM cancellation is not as effective as in the leading HQE term, 
operators of dimension 9 and dimension 12, see Fig.~(\ref{fig:D=9}),  might be numerically dominant.
In order to obtain an imaginary part of the loop integral, the operators of dimension 9 
have to be dressed with at least one gluon and the operators of dimension 12 with at least two gluons.
If we normalize the leading term (left figure of Fig.~(\ref{fig:D=6})) to 1, we
expect the D=9 diagram of Fig.~(\ref{fig:D=9}) to be of the order
${\cal O } (\alpha_s (4 \pi) \langle \bar{q} q \rangle   /m_c^3) \approx 0.03$
and the  D=12 diagram of Fig.~(\ref{fig:D=9}) to be of the order
${\cal O } (\alpha_s^2 (4 \pi)^2 \langle \bar{q} q \rangle^2   /m_c^6) \approx 10^{-3}$.
As explained above, the formally leading term of D=6 is strongly GIM suppressed to a value of
about $2 \cdot 10^{-5}$ and the big question is now how severe are the GIM cancellations
in the D=9,12 contribution. For the contributions to $y_D$ we get the naive expectations
\begin{displaymath}
\begin{array}{|c||c|c|}
\hline
y_D    & \mbox{no GIM}      & \mbox{with GIM}
\\
\hline
\hline
D=6,7  &    2 \cdot 10^{-2} &   5 \cdot 10^{-7}
\\
\hline
D=9  &      5 \cdot 10^{-4} &    ?
\\
\hline
D=12  &     2 \cdot 10^{-5} &    ?
\\
\hline
\end{array}
\end{displaymath}
If there would be no GIM cancellations in the higher OPE terms, then the D=9 or D=12 contributions
could be orders of magnitudes larger than the D=6 term, but in order to explain the experimental
number still an additional  numerical enhancement factor of about 15 has to be present.
For more substantiated statements $\Gamma_{12}^{D=9,12}$ has to be determined explicitly,
which is beyond the scope of this work \cite{D=9}.
This calculation is also necessary in order to clarify to what extent the large phase
in $\Gamma_{12}$ from the first OPE will survive. In order to determine the possible SM ranges of the 
physical phase $\phi$, in addition one has to determine $M_{12}$.
%
%
\phantomsection
\addcontentsline{toc}{subsubsection}{New physics}
\paragraph{New physics. ---}
Finally we would like to address the question, whether new physics (NP) can enhance $\Gamma_{12}$.
In the $B_s$ system it is argued \cite{Grossman:1996era} that $\Gamma_{12}$ is due to real intermediate
states, so one cannot have sizeable NP contributions. Moreover, the mixing phase
in the $B_s$ is close to zero, so the cosine in Eq.~(\ref{approx}) is close to one and therefore
NP can at most modify $\phi$, which results in lowering the value of $\Delta \Gamma$ compared 
to the SM prediction. In principle there is a loophole in the above argument. 
To $\Gamma_{12}$ also $\Delta B =1$ penguin operators contribute, whose Wilson coefficients 
might be modified by NP effects. But these effects would also change all tree-level $B$ decays. Since this 
is not observed at a significant scale, it is safe to say that within the hadronic uncertainties
$\Gamma_{12} = \Gamma_{12}^{SM}$ and therefore the argument of \cite{Grossman:1996era} holds.
Since in the $D^0$ system the QCD uncertainties are much larger, also the possible effects might be larger
but not dramatic. 
The peculiarity of the $D^0$ system -- the leading term in the HQE is strongly suppressed due to GIM 
cancellation -- gives us however a possibility to enhance $\Gamma_{12}$ by a large factor,
if we manage to soften the GIM cancellation. This might be accomodated either by
weakening the SU(3)$_F$ suppression in the first two terms of Eq.~(\ref{cancel}),
see, e.g., Petrov et al. \cite{Golowich:2006gq}, or by enhancing the CKM factors of the last two terms in
Eq.~(\ref{cancel}).
The latter can be realized in a model with an additional 
fourth fermion family (SM4). The usual CKM matrix is replaced by a four dimensional one
($V_{CKM4}$) and the unitary 
condition now reads $\lambda_d + \lambda_s + \lambda_b + \lambda_{b'} = 0$. Eq.~(\ref{cancel}) is replaced by
\begin{eqnarray}
\Gamma_{12} & = & - \lambda_s^2 \left( \Gamma_{ss}\! - 2 \Gamma_{sd} + \Gamma_{dd}  \right)
\label{cancel2}
\\ &&
              + 2 \lambda_s (\lambda_b + {\lambda_{b'}}) \left( \Gamma_{sd} - \Gamma_{dd} \right)
              - (\lambda_b + \lambda_{b'})^2 \Gamma_{dd} \;.
\nonumber
\end{eqnarray}
In \cite{Bobrowski:2009ng} an exploratory study of the allowed parameter space of $V_{CKM4}$ was performed
and as expected only very small modifications of $\lambda_d$ and $\lambda_s$ are experimentally allowed.
In almost all physical applications these modifications are numerically much smaller than the
corresponding hadronic uncertainties and therefore invisible. However, in 
the $D^0$ mixing system it might happen, that all dominant contributions cancel and only these modifications survive.
In the SM the first two terms of Eq.(\ref{cancel}) are numerical equal. In the SM4 the numerical 
hierarchy depends on the possible size of $\lambda_{b'}$, see Eq.(\ref{cancel2}). In particular, it was found
in \cite{Bobrowski:2009ng} that currently a value of $\lambda_{b'}$ of the order $\lambda^3$ is not excluded.
This means that the second term of Eq.(\ref{cancel2}) could be greatly enhanced by the existence 
of a fourth family and also the third term would now become relevant.
Using experimentally allowed data points for $V_{CKM4}$ from \cite{Bobrowski:2009ng} 
we have determined the possible values of $\Gamma_{12}$ in the SM4 from Eq.~(\ref{cancel2});
enhancement factors of a few tens are easily possible, see Fig.(\ref{fig:4gen}).
%
%
\phantomsection
\addcontentsline{toc}{subsubsection}{Discussion and Conclusions.}
\paragraph{Discussion and Conclusions. ---}
In this work we investigated the leading HQE contribution to the absorptive part of
$D$ mixing and corrections to it. We found that the size of these corrections is large, 
but not dramatic  ($\approx 50\%$ QCD, $\approx 30\% \, \, 1/m_c$). So it seems that the HQE might 
be appropriate to estimate the order of magnitude of $\Gamma_{12}$. We have explained that
$\Gamma_{12}^{D=6,7}$ gives, due to huge GIM cancellations, a value of $y$ which is about 
a factor of about 10000 smaller than the current experimental expectation, but it can have 
a large phase and it also does not vanish in the exact SU(3)$_F$ limit. 
Due to this peculiarity, it might be possible that the HQE result
is dominated by D=9 and D=12 contributions, if there the GIM cancellations are less pronounced, 
but to make more profound statements -- in particular about the standard model value of the
mixing phase -- these higher dimensional corrections have to be determined explicitly.
Finally we have shown that new physics can enhance $\Gamma_{12}$ by a high double-digit factor.
%
\paragraph*{Acknowledgment. ---}
\begin{acknowledgments}
We are grateful to I. Bigi and V. Braun for clarifying discussions.
\end{acknowledgments}
\providecommand{\href}[2]{#2}\begingroup\raggedright
\endgroup

\end{document}